\documentstyle{article}
\begin{document}
\begin{flushright}
{BI-TP 93/53}
\end{flushright}
\begin{flushright}
{Oktober 1993}
\end{flushright}
\begin{center}
{\bf RESUMMED EFFECTIVE ACTION IN INHOMOGENEOUS
EXTERNAL FIELD AT ZERO AND FINITE TEMPERATURE}\footnote{To appear in the
Proc. Workhsop  "Quantum Field Theoretical Aspects of High
Energy Physics" (Kyffh$\ddot{a}$user near Bad Frankenhausen, Germany, September
1993)}
\end{center}
\begin{center}
{\bf Andrei Leonidov\footnote{Alexander von Humboldt Fellow}}
\end{center}
\begin{center}
{\it Physics Department\\ University of Bielefeld D-4800,
Bielefeld,Germany\\ and \\Theoretical Physics Department\\ P.N.Lebedev
Physics Institute, 117924 Leninsky pr.53, Moscow, Russia}\\
\end{center}

\begin{abstract}
The two ways of resumming the effective action for the massless test
particles
in inhomogeneous external field at
zero and finite temperature providing the infrared finite answer are
discussed. The case of the massive test particles having a mass which is
parametrically small with respect to a scale set by the inhomogeneous
external field is briefly considered.
\end{abstract}

\newpage
\begin{center}
{\bf RESUMMED EFFECTIVE ACTION IN INHOMOGENEOUS
EXTERNAL FIELD AT ZERO AND FINITE TEMPERATURE}
\end{center}
\begin{center}
{\bf Andrei Leonidov}
\end{center}
\begin{center}
{\it Physics Department\\ University of Bielefeld D-4800,
Bielefeld,Germany\\ and \\Theoretical Physics Department\\ P.N.Lebedev
Physics Institute, 117924 Leninsky pr.53, Moscow, Russia}\\
\end{center}

\begin{abstract}
{\small The two ways of resumming the effective action for the massless test
particles
in inhomogeneous external field at
zero and finite temperature providing the infrared finite answer are
discussed. The case of the massive test particles having a mass which is
parametrically small with respect to a scale set by the inhomogeneous
external field is briefly considered.}
\end{abstract}
\vspace{5mm}
{\bf 1.} Below we discuss the ways of operationally defining the contributions
to
the effective action induced by interaction of a massless test scalar
particle with the external field. Namely, we shall be interested in
dealing with a well-known problem of the appearence of the
infrared-divergent contributions to the effective action when using the
expansion in the powers of the external field. The presentation follows
the two papers written in collaboration with A. Zelnikov [1,2]. In
this contribution we shall also discuss the small-mass expansion in the
case when a mass of the test particle is parametrically small with
respect to a characteristic scale set by an inhomogeneous external field
configuration.

Let us begin by reminding which problem we are going to deal with and
consider the simplest case of a scalar field $\varphi(x)$, interacting
with some background fields so that a corresponding Lagrangian has the
form
\begin{equation}
L[\varphi,V]={1 \over 2}\varphi(x)(-\Box+m^2+V(x))\varphi(x)
\end{equation}
where $\Box$ is a free Laplacian, $V$ is a potential depending on some
inhomogeneous external fields and the corresponding  effective action
 induced by
the test particles $\varphi(x)$ (or using the terminology from
statistical mechanics, the induced entropy of the external field) at the
one loop level is
\begin{equation}
W[V]=-{1 \over 2}\int_{0}^{\infty}{ds \over s} e^{-m^2 s}Tr(e^{-(-\Box+
V(x))s}-e^{\Box s})
\end{equation}
where we have used the proper time representation and the interesting
contribution is isolated by substracting the free
 propagation contribution.
In a situation when the external field potential
$V(x)$ is essentially inhomogeneous one is usually confined to using
only a certain number of terms in the expansion of the trace of the heat kernel
\begin{equation}
TrK(s)=Tr e^{-(-\Box+V(x))s}
\end{equation}
in the powers of the proper time $s$
\begin{equation}
TrK(s)=\sum_{n=0}^{\infty}b_{-\omega+n}(V)s^{-\omega+n}
\end{equation}
where $2\omega$ is a (eucledian) space-time dimension. We see that in the
massless case $m=0$ starting from some term in the heat kernel expansion
the corresponding effective action is term by term divergent at the
upper limit of the integration over $s$ and therefore the effective action
is an operationally ill-defined quantity in the infrared domain. In the
case when $m$ is parametrically small, the contribution of the higher
order terms to the effective action is proportional to the increasing
inverse powers of $m$, so that although these are formally infrared
finite, the expansion itself has no sence.
 The same problem arises in the finite temperature calculations of the
free energy of a system of massless particles in the inhomogeneous
static external field. Namely, we have for the free energy of this system at
the one-loop level
\begin{equation}
-\beta F={1 \over 2}\int_{0}^{\infty}{ds \over s}(1+2\sum_{n=1}^{\infty}
e^{-{4\pi^{2} s \over \beta^{2}} n^2} Tr K_{2\omega-1}(s))
\end{equation}
where $\beta$ is an inverse temperature
(let us recall that here the eucledian time is
compactified on a circle of a length $\beta$). We see that for $n=0$
(entropy) term when using a usual semiclassical expansinon for the spatial heat
kernel $K_{2\omega-1}(s)$ we face the same problem  of a term by
term infrared divergence at the upper limit of the integration over $s$.

As it is clear that the origin of a problem is in  expanding the
exponential in (3), the reasonable way to solve it is to make a
resummation of the proper time expansion keeping its exponential
character and thus providing the infrared convergent answer.
In the following we
shall consider two such approximations.

The first method corresponds to a summation of some sequence of terms of
all orders in the external field potential $V(x)$ using the analogy between
the trace of the heat kernel and the partition function, for which the
corresponidng resummation has been discussed in the literature [3]. We find
that for the simple external scalar field configurations the infrared
problem is cured and it is possible to get an infrared-finite answer for
the effective action (entropy) already for the simplest of previosly
discussed resummation for the partition function. The reasonable small
mass expansion is also easily obtainable.

The second  method corresponds to summing all derivatives for the terms
having some definite power in the external field. This corresponds to a
calculation of a nonlocal effective action [4-6,2]. In this method it was
proved
both at zero [4-6] and finite temperature [2] that this procedure allows to get
an
infrared-finite answer for the effective action (free energy). Below  we
illustrate this situation - again using the simplest localized scalar
field configurations and present some more general expressions for the
case of a finite temperature [2].

Let us notice, that formally it is possible to exponentiate the nonlocal
terms in the effective action too (at least for scalar [7] and electromagnetic
[8]
interactions). However it is not clear whether this method could
be used for actual computations.

 Let us also mention that in estimating the effective action one can use
the method of term-by-term infrared regularization of the effective
action using the series (4) and optimizing with respect to a cutoff [9],
but in this case it seems to be difficult to trace the interrelations
between the relevant scales to motivate the expansion parameter.
\vspace{5mm}

{\bf 2.} The first possible way of preserving the exponential character of the
heat kernel expansion is to keep the interaction potential $V$ in the
exponent as if it is constant, perform the derivative expansion and then
integrate over the space-time coordinates. This procedure is completely
analagous to the one used in statistical physics [3],
 where the analogue
of the trace of the heat kernel is a partition function, the inverse
temperature
being the analogue of the proper time. Let us now consider a simplest
case of a scalar interaction potential $V$. Then
in our notatons  the expression for the trace of the heat kernel for the
scalar particle propagating in the external scalar potential $ V(x)$ reads
$$
Tr(K(s)-K_{0}(s))=
$$
\begin{equation}
{e^{-m^2s} \over (4\pi s)^{\omega}} \int d^{2\omega} x
(e^{-V(x)s}(1-s^2 ({1 \over 6}\Box V(x)
-{1 \over 12} (\nabla V(x))^2+...)-1)
\end{equation}

Let us now consider a localized spherically symmetric static external field
configuration in $3$ dimensions correspoding to a scalar external field
potential
\begin{equation}
V(r) = {\Phi_{0}^2 \over (1+r/a)^{4}}
\end{equation}
and first calculate the effective action in the zero temperature
case. Neglecting for the moment the derivative terms in (6), we get
$$
W[\Phi_{0},a,m,M]=
$$
$$
-{1 \over 64\pi^2}\tau ({4 \over 3}\pi a^3)
(\Phi_{0}^2+m^2)^2
Log{M^2 \over \Phi_{0}^2+m^2} (-1+{4\Phi_{0}^2 \over \Phi_{0}^2+m^2}
-{104 \over 35}{\Phi_{0}^4 \over (\Phi_{0}^2+m^2)^2})
$$
\begin{equation}
-{1 \over 64\pi^2}\tau ({4 \over 3}\pi a^3)m^4
Log{M^2 \over m^2}
\end{equation}
where $\tau$ is an eucledian time and $M$ is an ultraviolet regulator. Let
us notice that at $\Phi_{0}=0$ the above expression vanishes. This
happens because we have substracted the free field contribution in (6).
Taking into account the derivative
terms  and keeping the lowest order term in the small mass
expansion, we get
\begin{equation}
W[\Phi_{0},a,M]=-{1 \over 64\pi^2}\tau ({4 \over 3}\pi a^3)\Phi_{0}^4
{1 \over 35}Log{M^2 \over \Phi_{0}^2} (1+{70 \over 63} {1 \over
\Phi_{0}^2 a^2}+{35 \over 2} {m^2 \over \Phi_{0}^2}) + ...
\end{equation}
We see that in the proposed approximation for the exponentiated
heat kernel the basic difference from the familiar constant external
field case (the first term in  the above expresion would just correspond to a
usual
effective potential) is the appearence of the effective volume $a^3$ of
the localized configuration instead of the total spatial volume
$V^{(3)}$. The logarithmic factor still depends only on the external
field amplitude  and taking into account the terms with derivatives
results in the expansion in the inverse powers of $\Phi_{0}^2 a^2$.

Let us now calculate the
free energy of a massless scalar field propagating in the static
backgorund field configuration Eq.7. We obtain
\begin{equation}
-({F \over T})={(\Phi_{0} a)^{3} \over 90}
-{2 \over 3} (\Phi_{0} a)^3 \sum_{n=1}^{\infty}(1+\mu n^2)^{3
\over 2}f(\mu,n)
\end{equation}
where $\mu={4\pi^2T^2 \over \Phi_{0}^2}$ and
\begin{eqnarray}
f(\mu,n^2) &=& ~_{2}F_{1}(1;-{3 \over 2};{3 \over 4};{1 \over 1+\mu n^2})
-~_{2}F_{1}(1;-{3 \over 2};{1 \over 2};{1 \over 1+\mu n^2}) \nonumber\\
&& +{1 \over 3}~_{2}F_{1}(1;-{3 \over 2};{1 \over 4};{1 \over 1+\mu
n^2}) - {1 \over 3} ({\mu n^2 \over 1+\mu n^2})^{3 \over 2}
\end {eqnarray}
where $~_{2}F_{1}(a;b;c;x)$ is a hypergeometric function and for
simplicity
we took into account only the leading exponential term (the
derivative corrections can be derived in a completely analogeous way).
{}From this expression one can work out the low- and high- temperature
expanisons in a standard way.
\vspace{5mm}

{\bf 3.} Let us now turn to the second possibility of constructing an
infrared-safe approximation for the calculation of the effective action
for massless test particles. This method corresponds to a summation of
all derivative terms for a given power of the external field. The
resulting effective action is therefore an essentially nonlocal object.
In a pioneering paper [4] Barvinsky and Vilkovisky have shown that this
procedure provides infrared-convergent integrals for the effective
action in all orders in the external field. Later this technique was
generalized to a finite-temperature case [2].
Below we shall analyse the explicitely nonlocal effective action for the same
case of a massless scalar field propagating in an localized external
field configuration [1] and consider the more general formal expressions for
the finite temperature case [2].

For the trace of the heat kernel we have a general expansion
\begin{equation}
TrK(s)=\sum_{n=0}^{\infty} TrK_{n}(s)
\end{equation}
where
\begin{equation}
TrK_{n}(s)={s^{n} \over n} \int_{\alpha_{i} \geq 0} d^{n}\alpha
\delta (1-\sum_{i=1}^{n} \alpha_{i})
Tr[Ve^{s\alpha_{1}\Box}...Ve^{s\alpha_{n}\Box}]
\end{equation}
and for the external field potential $V$ we shall take an
$O(3)$-symmetric external field configuration
\begin{equation}
V=\Phi_{0}^2 e^{-{r^2 \over 2a^2}}
\end{equation}
where the choice of the configuration to consider is dictated by
a computational simplicity. For the effective action at zero temperature
we get in the third order in the external field and keeping the lowest
order term in the small mass expansion
perturbation:
\begin{equation}
W={1 \over 64\pi^{{1 \over 2}}} {\tau \over a}(\Phi_{0}^2 a^2)^2 Log({1
\over M^2 a^2})[1+m^2a^2+...] - {c \over 192 (2\pi)^{1 \over 2}}({\tau \over a}
(\Phi_{0}^2 a^2)^3)
\end{equation}
where the constant $c=1.30348$ was obtained by numerical integration. We see
that the basic difference of this answer from that obtained in
the first section is that the charge renormalization logarithm is now
saturated by the slope of the field, and not by its amplitude and that
the expansion is now in powers of $\Phi_{0}a$. For the
free energy one gets in the same limit ($m=0$)
$$
-\beta F  =  {\pi \over 16}(\Phi_{0} a)^4
$$
\begin{equation}
 +{\pi^{{5 \over 2}} \over 4}
(\Phi_{0} a)^4 (aT)^2
\sum_{n=1}^{\infty} n^2 \int_{0}^{1}d\alpha\alpha^{-{3 \over 2}}
(1-\alpha)^{-{3 \over 2}}\Psi({3 \over 2},2,{4\pi^2a^2T^2n^2 \over
 \alpha(1-\alpha)})
\end{equation}
where $\Psi(a,c;x)$ is a confluent hypergeometric function. This
expression can serve as a starting point for constructing the low- and
high- temperature expansions by standard methods.

Let us now consider a more general case when we have external scalar and
Yang-Mills fields and develop an expression for the second order
effective action in the second order in the external fields at finite
temperature using the method of derivatives resummation [2]. Due to the
lack of space we shall not present here the details of the computation
and just present the final formulas demonstrating the above-described
derivative resummation providing the infrared-finite answer.

The formula for the trace of the heat kernel considered up to the second
order in the external field has a form
$$
Tr\ K(s)={1 \over (4\pi s)^{\omega}} \int d^{2\omega-1} x
\sum_{n=-\infty}^{\infty} tr[I+s\Phi -2s\Gamma_{0}^2\beta^2
 {\partial \over \partial \beta^2}
 $$
\begin{equation}
+s^2(\Phi f_4 (-s\Box) \Phi+G_{\mu \nu}f_5(-s\Box) G_{\mu \nu}
+4G_{\mu 0}f_5(-s\Box) G_{\mu 0}\beta^2 {\partial \over \partial
\beta^2})]exp(-{\beta^2 \over 4s}n^2)
\end{equation}
where $\Phi$ is an external scalar field, $G_{\mu \nu}$ is an
external Yang-Mills field, $\Gamma_{0}=\nabla_{\mu} (1/\Box) G_{\mu 0}$ and the
formfactors f
are the same as introduced in [6]:
\begin{equation}
f_{4}=\frac{1}{2}f, f_{5}=-\frac{1}{2} {f-1 \over 2},
f(-s\Box) =\int_{0}^{1} d\alpha exp(\alpha(1-\alpha)s\Box)
\end{equation}
The corresponidng $\zeta$-regularized expression for the effective
action in four space-time dimensions reads [2]
$$
W_{2}=-\int d^{2\omega-1} x \sum_{n=-\infty}^{\infty} \int_{0}^{1} d
\alpha
$$
$$
tr[{1 \over 4(4\pi)^2} \Phi[(n^2+B)^{\frac{1}{2}}-|n|]|\Phi
+{1 \over 2\beta^2}G_{\mu \nu}[(n^2+B)^{\frac{1}{2}}-|n|-{B \over 2|n|}]
G_{\mu \nu}
$$
\begin{equation}
+\beta^2 {\partial \over \partial (\beta^2)}{1 \over \beta^2}
G_{\mu 0}[(n^2+B)^{\frac{1}{2}}-|n|-{B \over 2|n|}]R_{\mu 0}]+
\Delta W_2
\end{equation}
where  the explicit expression for the local contribution for the
efective action $\Delta W_2$ arising due to the necessity of regularizing the
Matzubara frequency series can be found in [2]. The answer for the third
order term will have the same character (with the more complicated
structure of the parametric integration). The exponential formfactors in
the trace of the heat kernel (13) sandwiched between the external fields
provide the infrared finiteness of the effective action.

\begin{center}
 \it{Aknowledgments}
\end{center}
The author is grateful to Prof. H. Satz for kind hospitality at the University
of Bielefeld and for A.v.Humboldt
Foundation for financial support.
His work was
partially supported by the Russian Fund for Fundamental Research, Grant
93-02-3815.

\begin{center}
{\it References}
\end{center}

1. A.V.Leonidov, A.I.Zelnikov {\it {Resumming the Effective Action}},
preprint BI-TP 93/38, July 1993, hep-th 9308035;

2  A.V.Leonidov, A.I.Zelnikov {\it{Phys.Lett}} {\bf{B276}} (1992),122;

3. M.L.Goldberger, E.N.Adams. {\it{Journ. Chem. Phys.}} {\bf{20}} (1952),
240;

4. A.O.Barvinsky, G.A.Vilkovisky {\it{Nucl.Phys}} {\bf{B282}} (1987),163;

5. A.O.Barvinsky, G.A.Vilkovisky {\it{Nucl. Phys}} {\bf{B333}} (1990),471;

6. A.O.Barvinsky, G.A.Vilkovisky {\it{Nucl.Phys}} {\bf{B333}} (1990),512;

7. Y.Fujiwara, T.A.Osborn and S.F.J. Wilk {\it{Phys.Rev.}} {\bf{A25}}
(1982),14;

8. A.O.Barvinsky, T.A.Osborn {\it{A Phase-Space Technique for the
Perturbation Expansion of Schr$\ddot{o}$dinger Propagators}}, preprint
 MANI-92-01, May 1992;

9. D.I.Diakonov, V.Yu.Petrov and A.V.Yung {\it{Phys.Lett.}}{\bf{B130}}
(1983), 385; {\it{Sov.J.Nucl.Phys.}}{\bf{39}} (1984), 150.

\end{document}